\documentclass[%
preprint,
11pt,
 amsmath,amssymb,
aps,
prf,
longbibliography
]{revtex4-2}

\usepackage{graphicx}
\usepackage{epstopdf}
\usepackage{dcolumn}
\usepackage{bm}
\usepackage{xcolor}
\usepackage{wrapfig}

\begin{document}


\title{A stochastic model for the residence time of solid particles in turbulent Rayleigh-B\'enard Flow}

\author{Colin J. Denzel}
\affiliation{Department of Aerospace and Mechanical Engineering, University of Notre Dame, Notre Dame, IN, USA}

\author{Andrew D. Bragg}%
\affiliation{%
Department of Civil and Environmental Engineering, Duke University, Durham, NC, USA
}%

\author{David H. Richter}%
\affiliation{%
Department of Civil and Environmental Engineering and Earth Sciences, University of Notre Dame, Notre Dame, IN, USA
}%

\date{\today}

\clearpage

\begin{abstract}

The Pi Chamber, located at Michigan Technological University, generates moist turbulent Rayleigh-B\'{e}nard flow in order to replicate steady-state cloud conditions. We take inspiration from this setup and consider a particle-laden, convectively-driven turbulent flow using direct numerical simulation (DNS). The aim of our study is to develop a simple stochastic model that can accurately describe the residence times of the particles in the flow, this time being determined by the complex competition between the gravitational settling of the particles, and the interaction of the particles with the turbulent structures in the flow. A simple conceptual picture underlies the stochastic model, namely that the particles take repeated trips between the top and bottom boundaries, driven by the convective cells that occur in Rayleigh-B\'{e}nard turbulence, and that their residence times are determined by the time it takes to complete one of these trips, which varies from one trip to another, and the probability of falling out to the bottom boundary after each trip. Despite the simplicity of the model, it yields quantitatively accurate predictions of the distribution of the particle residence times in the flow. We independently vary the Stokes numbers and settling velocities in order to shed light on the independent roles that gravity and inertia play in governing these residence times. 

\end{abstract}

\maketitle

\section{Introduction}
\label{intro}
\vspace{-12pt}
The settling of inertial particles in turbulent flows is relevant to a wide array of engineered and natural systems, including the dispersion of pollutants \citep{SchwaigerAsh}, the settling of organic materials in the ocean \citep{RuizPlankton}, and the cooling of Earth's magma \citep{PatockaMagma}. In this study we are particularly motivated by experiments conducted in the so-called ``Pi Chamber'', a cloud chamber facility located at Michigan Technological University which uses two temperature-controlled, saturated plates in order to replicate cloud conditions via moist turbulent Rayleigh-B\'{e}nard (RB) flow. The chamber itself has been described extensively elsewhere \citep{ChangPi}, including efforts to characterize unladen RB flow in moist conditions \citep{ChandrakarJFM2020}, and for this work it serves as a broad motivation for understanding the Lagrangian dynamics of particles, especially their gravitational sedimentation. To this end, the Pi Chamber will serve as the starting point, although our exploration will extend beyond the properties of the particles seen in the experimental facility itself; i.e., our analysis spans ranges of non-dimensional parameters that cannot be replicated experimentally in order to gain further insight into the separate roles of gravity and inertia on the particle residence times in the flow.

Existing studies on particle laden RB turbulence have largely focused on how thermal and dynamic coupling affects turbulence and particle motion, primarily via two-way coupled simulations that attempt to take into account all of the physics relevant to the onset of turbulence and the transfer of heat and momentum \citep{Oresta2014, Lakkaraju2013, Scanlon1973}. In a somewhat similar setup to the Pi Chamber, \citet{Oresta2013} simulated RB flow and allowed solid, isothermal particles to settle from the top boundary. Over a wide range of particle diameters, they found that mechanical and thermal coupling were able to change the mean particle settling velocities. The results also suggested a tendency towards ``reverse one-way coupling'' where varying fluid parameters had a small effect on the behavior of the particles. In light of this, we are interested in particle residence times as a function of particle properties and will only consider one set of flow parameters, with the understanding that for a sufficiently turbulent environment, all of the relevant mechanisms will be present. 

The study of isothermal, inertial particles was furthered in \citet{Yang2022}, which considered particles with three different Stokes numbers ($St$). They found that both heat and momentum transfer were significantly enhanced for the medium Stokes number due to strong coupling of the two phases, while the coupling for the lowest and highest $St$ was weak. This non-monotonic relationship between particle dispersion and $St$ has been frequently observed in other particle-laden turbulent flows \citep{Brandt2022}. In an effort to isolate the effects of inertia from gravitational forces in turbulent, two-way coupled RB flow, \citet{Park2018} looked at non-isothermal particles and varied $St$ and a scaled settling velocity ($Sv$) independently, allowing for a more detailed exploration of momentum coupling. Although this study was more focused on how thermal and mechanical coupling changed the turbulent kinetic energy (TKE) and Nusselt number ($Nu$) of the flow, the approach of independently varying $St$ and $Sv$ is an essential component of the current work. 

The discussion of settling rates and residence times of small heavy particles is a well-studied aspect of turbulent flows in general. Historically, these efforts have been focused on isotropic, homogeneous turbulence with zero mean velocity. It has been demonstrated that the settling of these particles is dependent on the particle inertia and the free-fall terminal velocity. When there is no inertia, the particles on average settle at the same rate as in still fluid. However, inertia can create a bias for particles to move towards downwards-sweeping regions of the flow \citep{Maxey1987}. The resulting mechanics have since been studied extensively \citep{Aliseda2002, Wang1993,tom_2019}, showing that inertial clustering and gravitational settling lead to preferential sweeping and ultimately increased settling velocities when compared to the velocity predicted by Stokes drag in a quiescent medium. In \citet{Rosa2016}, where it was shown that preferential sweeping was the dominant means of increasing average settling velocity, the inertial and gravitational settling parameters were separated by varying the ratio of particle to fluid density and the energy dissipation rate. Furthermore, another mechanism proposed in \citep{Rosa2016} is called loitering, which refers to falling particles spending more time in regions with upward flow, ultimately reducing average settling velocities. In numerical simulations of homogeneous isotropic turbulence, this mechanism only plays a role when non-linear drag is considered \citep{Good2014}, which is not used in the present simulations. 

In the present setup, we also must account for the effects of boundary layers near the wall. The settling velocities of inertial particles through wall-bounded turbulence was studied in \citet{Bragg2021,Bragg2021_2}, which explored theoretically the physical mechanisms governing the particle transport, and used DNS to explore how the various mechanisms contribute as $Sv$ and $St$ are varied. While it was evident that the well-known effects of preferential sweeping were present in the bulk of the flow, the contribution of this mechanism decreased near the wall. In the near-wall region where the gradients in the turbulence statistics are strong, the turbophoretic drift mechanism takes over and becomes the dominant mechanism responsible for the enhanced settling speed of the particles. This is in fact the same mechanism that is also responsible for a build up of the particle concentration near the wall even in the absence of settling \citep{Reeks1983}. These additional considerations complicate the problem and has led to the implementation of stochastic models of varying complexity. 

Understanding and modeling particle settling rates and residence times are relevant to understanding the formation of cloud droplets in the Pi Chamber. In \citet{Saito2019}, the focus was on the development of a Fokker-Planck equation and its prediction of the broadening of droplet size distributions (DSD) in experimental clouds. In order to facilitate the comparison with statistical theory, they took the simplest approach by assuming that the removal process (and therefore droplet lifetime) was independent of particle size. In a similar exploration of an evolving Pi Chamber DSD, \citet{Krueger2020} assumed that when a droplet becomes sufficiently close to the lower boundary, the probability of falling out per unit time is determined by the terminal velocity, which is assumed to follow Stokes law and is therefore proportional to the square of the radius. In this study, we show that these assumptions \citep{Saito2019, Krueger2020} are valid, but only within certain regimes of $St$ and $Sv$. Our proposed model accounts for the relevance of the B\'{e}nard cells and allows us to separate the effects of inertia and gravitational forces. This is achieved through a small number of parameters relevant to RB flow that are dependent on $St$ and $Sv$. We will demonstrate how these parameters vary with particle properties, and how these variations ultimately determine residence times. 

\clearpage

\section{Methods}
\label{methods}

\subsection{Numerical Setup}
To generate statistical data for the development of our stochastic model, we employ a direct numerical simulation (DNS) of the turbulent Rayleigh-B\'{e}nard flow, which is then one-way coupled with Lagrangian particles, allowing for the particle statistics to be recorded in a Lagrangian frame of reference. As noted above, the setup is broadly motivated by the conditions found in the Pi Chamber, and therefore is similar to the methods used by \citet{MacmillanPi}, except that in the present case the particles are non-evaporating and one-way coupled to the surrounding flow. While we will provide a brief overview of the DNS model as it pertains to this study, further details can be found in \citet{RichterFog}, \citet{Park2018}, and \citet{Helgans2016}. 

The Navier-Stokes equations with the Boussinesq approximation are solved for mass, momentum, and energy conservation of the carrier phase:
\begin{equation} \label{eq:mass}
    \nabla \cdot \mathbf{u} = 0,
\end{equation}
\begin{equation} \label{eq:momentum}
   \frac{\partial \mathbf{u}}{\partial t}+\mathbf{u} \cdot \nabla \mathbf{u}= -\nabla \pi+\hat{\mathbf{k}} \frac{g}{T_0} T + \nu \nabla^2 \mathbf{u},
\end{equation}
\begin{equation} \label{eq:energy}
    \frac{\partial T}{\partial t}+\mathbf{u} \cdot \nabla T = \alpha \nabla^2 T,
\end{equation}
where $\mathbf{u}$ is the fluid velocity, $T$ is the temperature, and $\pi$ is a pressure variable which enforces the divergence-free condition of Eq. \ref{eq:mass}. In Eq. \ref{eq:momentum}, the buoyancy term in the vertical direction is dependent on the acceleration due to gravity $\mathbf{g}=g\hat{\mathbf{k}}$, and the reference temperature $T_0 = 300 K$. The terms $\mathbf{\nu}$ and $\mathbf{\alpha}$ refer to the kinematic viscosity and heat diffusivity of the fluid. Since we are considering solid, isothermal, one-way coupled particles, there is no need for the additional source terms from particle coupling that are found in \citet{MacmillanPi} and \citet{Park2018}; this will allow us to vary the gravity felt by each particle in later analysis without concern for the effects that the particles may have on one-another or the flow. 

Along the upper and lower boundaries, the fluid velocity is governed by a no slip condition. The aspect ratio of the domain is $L_{x}/L_{z} = L_{y}/L_{z} = 2$, and the number of grid points is $[N_x, N_y, N_z] = [128, 128, 128]$. The 2:1 aspect ratio similar to that found in the Pi Chamber. However, unlike the Pi Chamber, the domain is horizontally periodic owing to the pseudospectral discretization in the $x$ and $y$ directions. Second order finite differences are employed for derivatives in the vertical $z$ direction. The temperature of the upper and lower boundaries of the rectangular domain are set to $T_{top}=280~\textrm{K}$ and $T_{bot}=299~\textrm{K}$, resulting in a temperature difference $\Delta T = 19~\textrm{K}$. This corresponds to a Rayleigh number of $Ra \equiv (g \Delta T L_z^3) / (T_{0} \nu \alpha) = 10^7$. This value is used for all simulations in this study, along with a Prandtl number $Pr \equiv \nu/ \alpha = 0.715$.

The particles evolve in a Lagrangian frame of reference according to the following set of equations: 
\begin{equation} \label{eq:xp}
    \frac{d \mathbf{x}^{i}_{p}}{d t}= \mathbf{v}^{i}_{p},
\end{equation}
\begin{equation} \label{eq:vp}
    \frac{d \mathbf{v}^{i}_{p}}{dt} = \frac{1}{\tau_{p}} \left( \mathbf{u}_{f} - \mathbf{v}^{i}_{p} \right) - g_{p} \mathbf{\hat{k}},
\end{equation}
where the evolution of the $i^{th}$ particle's position $\mathbf{x}_{p}^{i}$ and velocity $\mathbf{v}_{p}^{i}$ depend solely on $\tau_p$, ${g}_p$, and the local fluid velocity $\mathbf{u}_{f}$ interpolated to the droplet location using trilinear interpolation. All particles begin their lifetime at the midplane with zero initial velocity and are taken out of the flow when they reach the bottom boundary. The timescale $\tau_{p} = \rho_a d^{2} / 18 \nu \rho_{f}$ is the Stokes timescale, which governs the time taken by a particle of diameter $d$ to reach equilibrium with the local velocity of a fluid that has density $\rho_f$ and kinematic viscosity $\nu$. The gravitational acceleration experienced by the particle, ${g_p}$, is separate from that experienced by the fluid ${g}$, thus allowing us to specify the particle settling rate independent of the buoyancy forcing of the fluid. Using Kolmogorov microscales to non-dimensionalize the velocities in Eq. \ref{eq:vp}, we obtain the following:
\begin{equation} \label{eq:vpnd}
    St\frac{d \mathbf{\tilde{v}}^{i}_{p}}{d\tilde{t_k}} = \left( \mathbf{\tilde{u}}_{f} - \mathbf{\tilde{v}}^{i}_{p} \right) - Sv~\mathbf{\hat{k}}.
\end{equation}
As a result, it is evident that particle motion is solely dependent on two non-dimensional parameters: Stokes number ($St)$ and settling velocity ($Sv$). For the purposes of this study, they will be defined as $St = \tau_{p} / \tau_{k}$, and $Sv = \tau_{p}g_{p} / v_{k}$ where $\tau_{k}$ and $v_{k}$ are the vertically averaged Kolmogorov time and velocity microscales, respectively. Note that the non-dimensional time, $\tilde{t_k} = t/\tau_K$ is used in Equation \ref{eq:vpnd} to define $St$ and $Sv$. A separate time scale that will be used extensively in this work is a convective time scale defined as $\tilde{t_c} = t/\tau_e$. The parameter $\tau_e$ is the eddy turnover time given by $\tau_e = 2L_z/\sqrt{\langle u_z^2 \rangle _{V,t}}$, where $\langle\rangle_{V,t}$ indicates a volumetric and temporal mean of the vertical velocity ($u_z$) squared; a turnover time definition used elsewhere in RB studies \citet{Sakievich2020}. For this simulation, $\tau_e = 3.95$ minutes; how this relates to predicting particle residence times will be explored in a later section. 

Unladen RB flow has been studied extensively in the literature \citep{Bodenschatz2000}, and therefore will not be the focus of extensive discussion in this study. More details on the specific setup used to inspire/generate the flow in this application can be found in \citet{ChangPi} and \citet{MacmillanPi}. The focus for the remainder of this paper will be on understanding and modeling the lifetime behavior of solid, isothermal particles as a function of $St$ and $Sv$. 

\subsection{Model Description}

The processes and mechanisms that govern particle transport in turbulent Rayleigh-B\'{e}nard flow are very complicated. However, the conceptual framework behind our model assumes that a relatively small number of flow quantities and simple processes determine the probability distribution of the particle residence times. This conceptual framework is inspired by the observation that particles are circulated globally by the convection cells in the flow, causing them to take a number of `elevator trips' before eventually falling out; the number of these trips largely influences the total residence time. 

Concerning the dependence of the particle dynamics on $St$ and $Sv$, a couple of limiting cases have straightforward interpretations. In the limit of zero inertia and terminal velocity, i.e,. $Sv \rightarrow 0$ and $St \rightarrow 0$, the particles will act as fluid tracers. These are continuously circulated by the convection with no chance of falling out in finite time because we do not consider diffusive processes. In contrast, as $Sv \rightarrow \infty$ and $St \rightarrow \infty$ the infinite inertia eliminates the effect of the flow and prevents the particle from ever accelerating to its terminal velocity. Another frequently made simplification assumes that $Sv \sim$ finite and $St \rightarrow 0$. In this case, for particles that are initially distributed homogeneously, the mean velocity of the particles would be the Stokes settling velocity because their lack of inertia means that the particles sample the flow uniformly for all times. This is the assumption behind the well-known Rouse profile of suspended particulate matter \citep{RouseASCE1937}. In the majority of applications, however, including the droplets found in the Pi Chamber, the presence of a finite non-zero $St$ introduces the complicating role of inertia. 

\begin{figure}[t]
    \centering
    \includegraphics[width=0.98\textwidth]{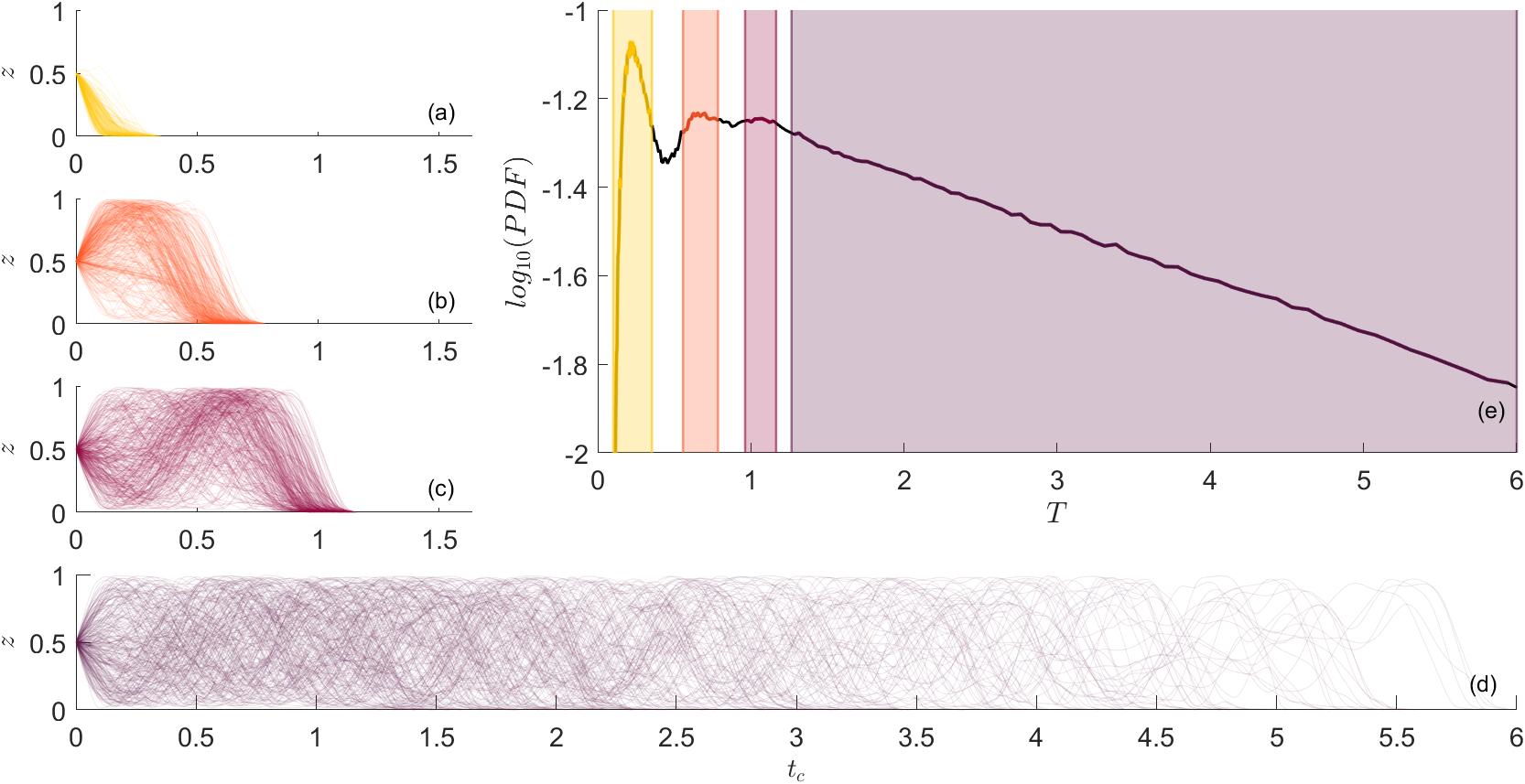}
    \caption{Sample distribution of the non-dimensional residence times ($T$) as defined in Eq. \ref{eq:gov}, resulting from the DNS (e). To emphasize the importance of elevator trips on the qualitative nature of the distribution, see the trajectories from the first (a), second (b), and third (c) peaks, as well as the tail of the distribution (d).}
    \label{fig:elev_example}
\end{figure}

The elevator trips that inspire this model are highlighted in Figure \ref{fig:elev_example}, which shows a sample probability distribution of particle residence times from the DNS non-dimensionalized by $\tau_e$ (Figure \ref{fig:elev_example}(e)), and identifies with color shading representative trajectories that correspond to its most obvious features. 
If the trajectories were approximated to be sinusoidal, we could consider one of these `elevator trips' to be one period that begins and ends at the midplane. The first two peaks in the distribution are a result of those that either had an initial downward velocity and only traversed the distance from midplane to bottom boundary, or those that had an initial upward velocity and completed one-half period before traversing the same final distance. The remaining particles then complete an unspecified number of elevator trips before ultimately traversing the distance from the midplane to the bottom boundary. It is then evident that, in order to approximate the residence time of a particle, we need four pieces of information: 
\begin{enumerate}
    \item ($\lambda_d$) How likely the particle is to have an initial downward velocity 
    \item ($\rho_e$) The amount of time it takes the particle to complete one elevator trip
    \item ($\lambda_f$) How likely the particle is to fall out of the flow after each trip 
    \item ($\rho_b$) How much time it takes the particle to traverse from the midplane to the bottom boundary before falling out 
\end{enumerate}
These are the four parameters that the stochastic model takes into account to predict the full distribution of residence times, and are expected to be a function of both $St$ and $Sv$. Our goal is to demonstrate this dependence in the formulation of this model. For the remainder of this work, the reported values of $\rho_e$ and $\rho_b$ will be non-dimensionalized by the eddy turnover time $\tau_e$. 

\begin{figure}[t]
    \centering
    \includegraphics[width=0.98\textwidth]{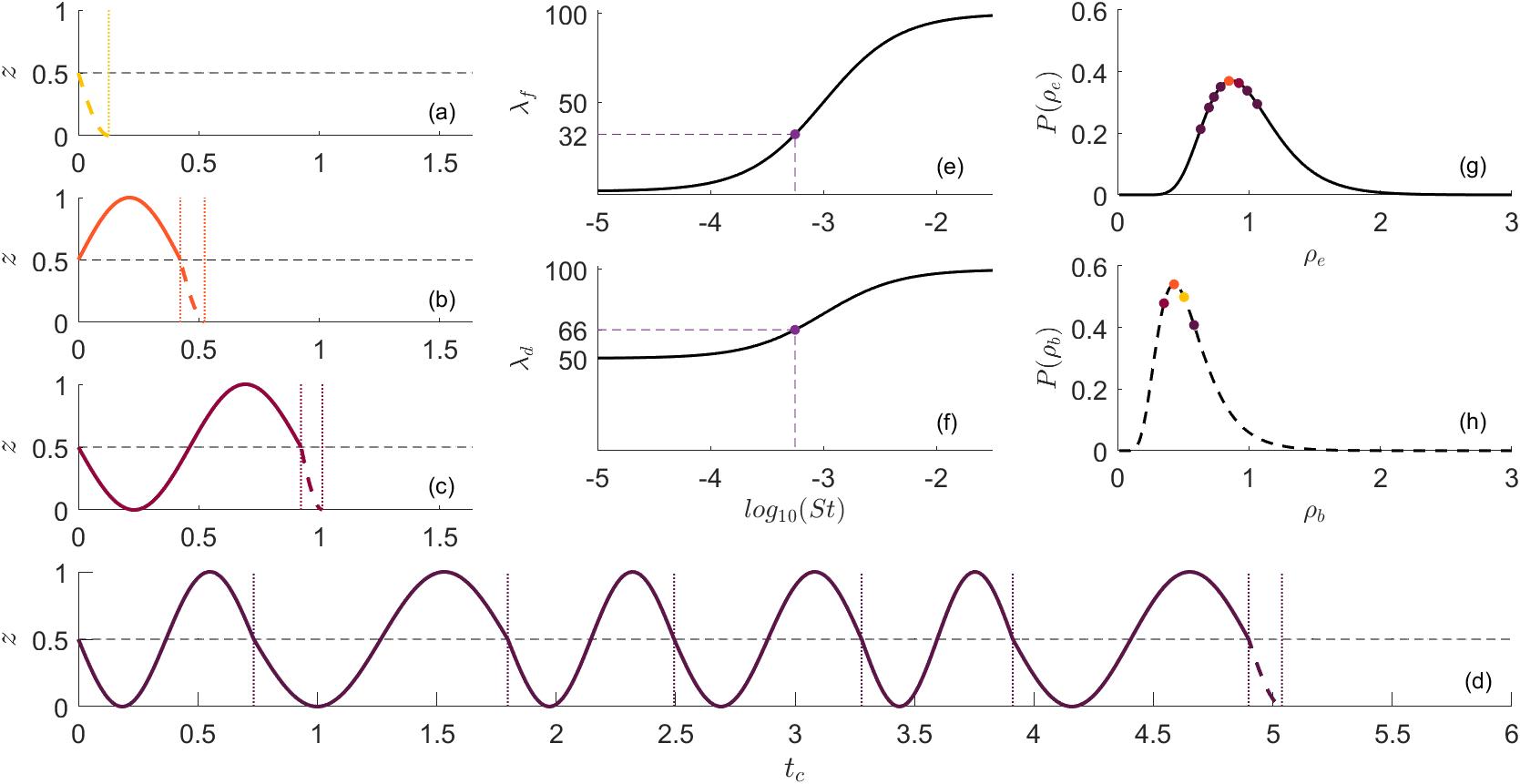}
    \caption{Demonstration of the stochastic model for four trajectories given particles of the same size, where $\{\widetilde{\xi} = -1,~N_t = 0\}$ (a), $\{\widetilde{\xi} = 1,~N_t = 0\}$ (b), $\{\widetilde{\xi} = -1,~N_t = 1\}$ (c), and $\{\widetilde{\xi} = -1,~N_t = 6\}$ (d). The residence time predictions take into account the chance of falling out after each elevator trip, $\lambda_f$ (e), the likelihood of having an initial downward velocity, $\lambda_d$ (f), the time required to complete an elevator trip, $\rho_e$, (g), and the time to traverse from the midplane to the bottom boundary, $\rho_b$ (h).}
    \label{fig:model_demo}
\end{figure}

In order to construct a stochastic model based on this conceptual framework, two steps are required to determine the residence time of the $i^{th}$ particle, $T_i$. Step 1: determine whether the particle has an initial upward or downward velocity based on $\lambda_d$. While no additional considerations must be taken for an initial downward velocity, if it is upward, add the time associated with half of an elevator trip $(\widetilde{\rho_e}/2)$. Step 2: determine if the particle will fall out of the flow based on $\lambda_f$. If it does, add the time required to pass from the midplane to bottom boundary $\widetilde{\rho_b}$ and consider the particle dead. If the particle does \textit{not} fall out, add the time required for an elevator trip $\widetilde{\rho_e}$, and repeat step 2 until the particle does fall out of the flow. How this simple process replicates the trajectories shown in Figure \ref{fig:elev_example} is demonstrated visually in Figure \ref{fig:model_demo}. In Figure \ref{fig:model_demo}(a,b), we see particles that complete zero elevator trips and have initial downward and upwards velocities respectively. Figure \ref{fig:model_demo}(c) shows a particle completing one elevator trip before falling out, while Figure \ref{fig:model_demo}(d) shows a particle completing multiple elevator trips. 

The procedure described above may be summarized mathematically as follows. Let $\widetilde{\xi}$ denote a random variable living in the discrete sample-space $\xi$ that takes values $-1$ and $+1$, with probability $\mathbb{P}(\widetilde{\xi}=-1)=\lambda_d$, and hence $\mathbb{P}(\widetilde{\xi}=+1)=1-\lambda_d$. The realization $\widetilde{\xi}=-1$ is used to denote that the initial particle velocity is down, while $\widetilde{\xi}=+1$ denotes that it is up. The non-dimensional residence time for the $i^{th}$ particle, $T_i$, is then specified by the model to be
\begin{align} \label{eq:gov}
T_i&=\sum_{j=0}^{N_t}\beta(j,\widetilde{\xi})\widetilde{\rho}_e^{(j)}+\widetilde{\rho}_b
\end{align}
where $\beta(j=0,\widetilde{\xi}=-1)=0$, $\beta(j=0,\widetilde{\xi}=+1)=1/2$, and $\beta(j,\widetilde{\xi})=1~\forall j>0,\widetilde{\xi}$, with $N_t$ being the total number of elevator trips taken by the particle before it falls out. In the expressions above, $\widetilde{\cdot}$ is used to denote the random variable conjugate to the sample-space variable, e.g., $\widetilde{\rho}_b$ is the random variable conjugate to ${\rho}_b$.

In order for this new model to be fully closed and predictive, the statistical quantities summarized above would have to be modeled. However, as a first step we simply specify them using the DNS data. The advantage of doing this is that it allows the simple conceptual idea underlying the stochastic model to be tested. Given the complexity of particle motion in turbulent Rayleigh-B\'{e}nard flow, it is not at all obvious \textit{a priori} that our simple conceptual framework is sufficiently detailed to quantitatively capture the particle residence times in the flow. Once the accuracy of this conceptual modeling framework has been established, it will then make sense to try to model the input statistics and so derive a fully closed, fully predictive model.

\clearpage

\section{Results}
Here we discuss the particle residence time behavior behavior observed in the DNS, as well as the performance of the model as $St$ and $Sv$ are varied. The varying of $St$ and $Sv$ follows two different strategies. In the first, $Sv$ and $St$ are inherently coupled, as they would be in a physical experiment where the acceleration of gravity felt by the particles $g_{p}$ would equal that responsible for the buoyancy forcing $g$. While this is consistent with physical experiments, it does not allow for the effects of gravity and inertia to be untangled, which can hinder an understanding of the problem. To explore this, we therefore also consider cases where $Sv$ is held constant (by varying $g_{p}$) while varying $St$, allowing us to distinguish the effects of gravitational settling from particle inertia on the particle residence times.

\subsection{Coupled $St$ and $Sv$}
For the coupled case, we consider particles with a range of Stokes numbers $St = [10^{-6},10^{-1}]$, which, since the flow and particles experience the same gravitational acceleration, implies the range $Sv = [10^{-3},10^{2}]$. For reference, a 0.5 micron salt aerosol in the Pi Chamber has $St \sim \mathcal{O}(10^{-6})$ and $Sv \sim \mathcal{O}(10^{-3})$, and a 20 micron cloud water droplet has $St \sim \mathcal{O}(10^{-3})$ and $Sv \sim \mathcal{O}(10^{0})$. While our range encompasses realistic values, we are also intentionally considering a wider range in order to more comprehensively understand the problem and test the model. 

\begin{figure}[t]
    \centering
    \includegraphics[width=0.98\textwidth]{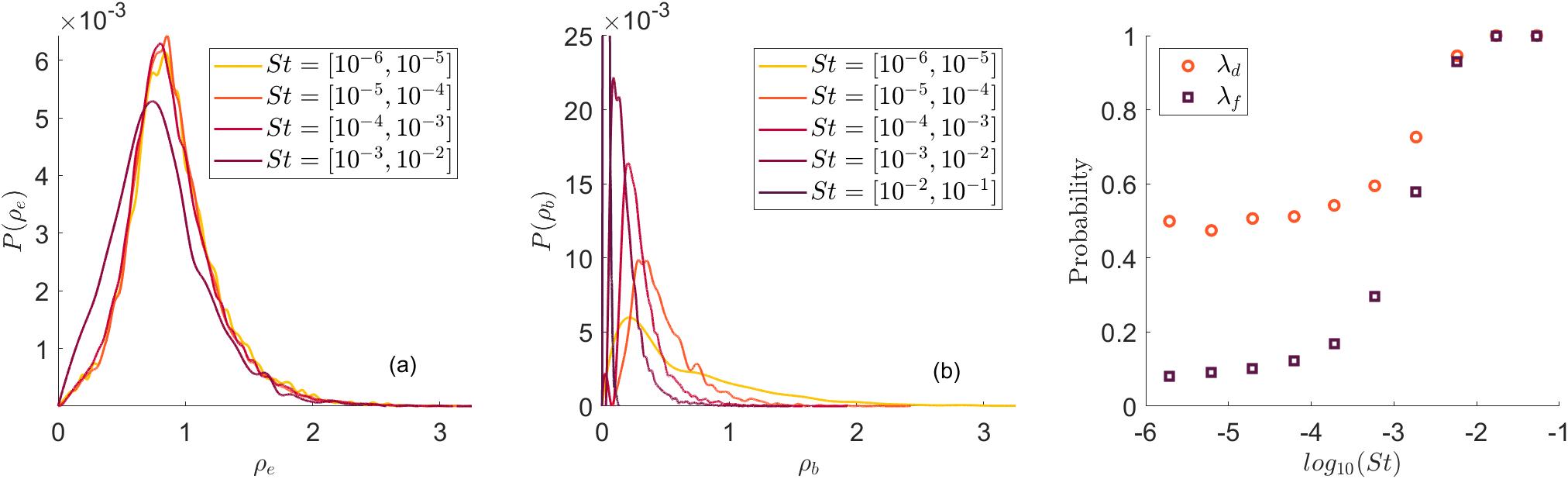}
    \caption{The four input parameters from the coupled $St$ and $Sv$ case, the distribution of $\rho_e$ (a): the distribution of $\rho_b$ (b), as well as $\lambda_d$ and $\lambda_f$ (c) expressed as a probability}
    \label{fig:coupled_stats}
\end{figure}

\begin{figure}[t]
    \centering
    \includegraphics[width=0.95\textwidth]{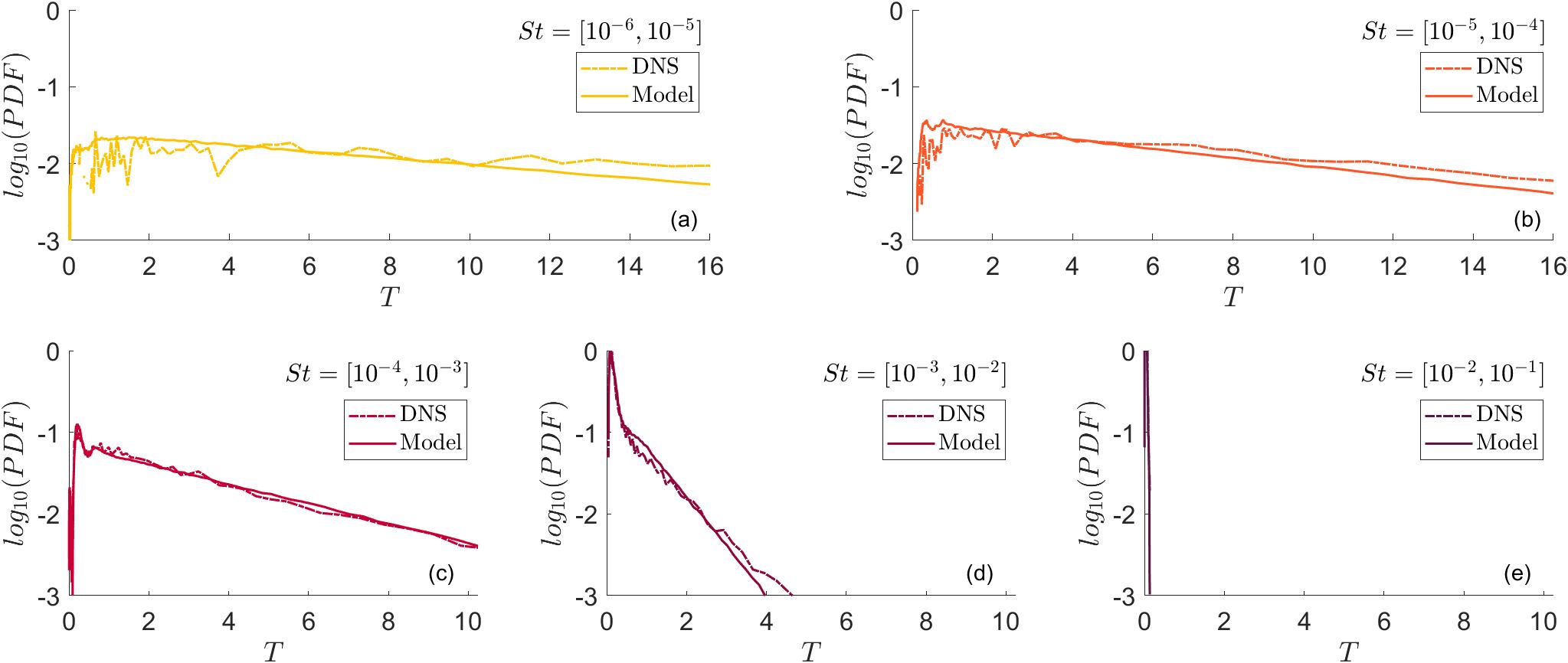}
    \caption{The results of the coupled St and Sv case as shown by the comparison of non-dimensional residence time ($T$) distributions for the DNS and the model. For clarity, each order of magnitude of St is compared individually (a-e)}
    \label{fig:coupled_pdfs}
\end{figure}

\begin{figure}[t]
    \centering
    \includegraphics[width=0.98\textwidth]{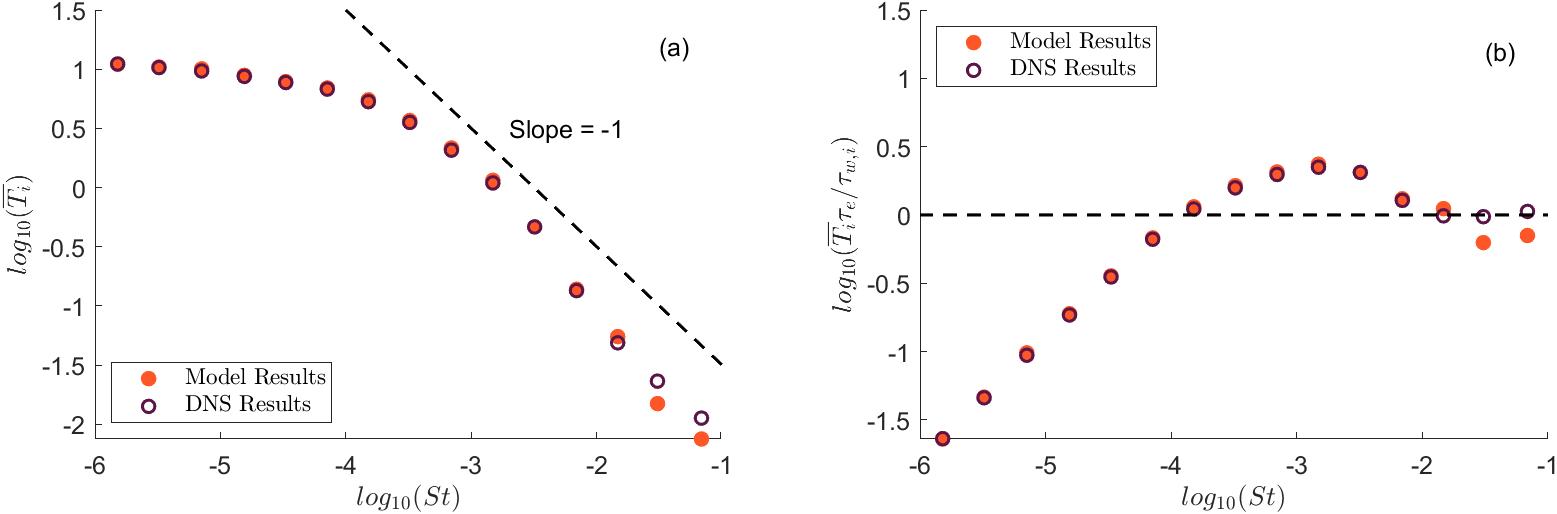}
    \caption{The results of the coupled St and Sv case as shown by the comparison of mean non-dimensional residence times ($\overline{T_i}$), with a reference slope included for comparison to the power law relationship predicted by Stokes drag (a). The same results are also plotted where the mean residence time is instead non-dimensionalized by the settling time of a given particle in quiescent flow ($\tau_{w,i}$) (b).}
    \label{fig:coupled_means}
\end{figure}

The four statistical quantities required for the model are shown in Figure \ref{fig:coupled_stats}, as measured by the DNS and used in the model results of Figures \ref{fig:coupled_pdfs} and \ref{fig:coupled_means}. For both distributions, a discrete PDF is generated directly from a the DNS data and then fit with cubic splines in order to create a continuous CDF. The model then uses inverse transform sampling by pulling from this CDF to generate pseudo-random numbers that adhere to the probability distribution of our choice. The PDF of period residence times, $P(\rho_e)$, seen in Figure \ref{fig:coupled_stats}(a), clearly shows that for this range of properties, the majority of particles complete an elevator trip in accordance with the convective time ($t_c$). Note that there is no data for the largest $St$, as at this size no particles completed an elevator trip. The distribution of times for a particle to traverse from the midplane to the bottom boundary, $P(\rho_b)$, seen in Figure \ref{fig:coupled_stats}(b), however, is strongly correlated with $St$, with smaller times being associated with larger particles. This is because the increased settling velocity and inertia of the particles leads to a larger terminal velocity and increasingly negligible effects of the flow. The chance of having an initial downward velocity, $\lambda_d$, is approximately 50\% for the smaller $St$, but rises to 100\% for the largest. Similarly, the percent chance of falling out during an elevator trip, $\lambda_f$, starts below 10\% but rises to 100\% for the larger $St$, meaning all particles of that size drop out as soon as they are injected into the flow.

With these four inputs, Figure \ref{fig:coupled_pdfs} compares the probability density functions (PDFs) of residence times as measured in the DNS to those generated by the model. Figure \ref{fig:coupled_means} compares the mean residence times predicted by DNS and the model as a function of $St$. We see that the model captures nearly all relevant features of the DNS PDF, including the peaks at low $t_c$, and is quantitatively accurate for all ranges of $St$. Unsurprisingly, we see that the larger particles, those with both high $St$ and $Sv$ values, on average fall out faster than their smaller counterparts. Recalling that the residence times in Figures \ref{fig:coupled_pdfs} and \ref{fig:coupled_means}(a) have been normalized by the convective time scale of the flow, we see the two expected peaks around 1/4 and 3/4 periods. Beyond those peaks, the linear nature of the log-scale PDFs suggest qualitative agreement with the results of \citet{PatockaMagma}, who demonstrated that the number of suspended particles in a system could be robustly modeled with an exponential decay relationship dependent on the settling velocity and flow properties. In Figure \ref{fig:coupled_means}, we see that for $St < 10^{-4}$, the particles all take an average of 10 elevator trips before falling out. In this regime, the assumption made in \citet{Saito2019} that lifetime is independent of particle properties may be sufficient. It appears that the trajectory is dominated by convection, and the slight non-zero slope is only caused by the different settling velocities as the particle traverses the boundary layer. These results are qualitatively similar to what was seen by \citet{PatockaMagma}, which also identified a regime of slow sedimentation dominated by large-scale circulation. The number of trips continues to decrease, until beyond $St = 10^{-3}$ we see that the residence time is on average less than one convective time scale. This decrease in residence times is a result of the higher chance of falling out after each elevator trip, and the decreased time to pass from the midplane to the bottom boundary. It is also in this regime that we see agreement with the simplification made in \citet{Krueger2020}, which assumed that settling rate follows Stokes Law and that the mean lifetime is therefore proportional to $1/St$ (shown by the reference slope in Figure \ref{fig:coupled_means}(a)). This trend is perhaps more obvious in Figure \ref{fig:coupled_means}(b). It shows the mean residence time non-dimensionalized by $\tau_w$, which is the amount of time a given particle will take to settle in quiescent flow, defined as $\tau_w = 0.5L_z/(\tau_p g_p)$. In this context, it is evident that particles with $St > 10^{-2}$ fall out at the rate predicted by Stokes Law. However, it is still unclear whether this transition is associated with inertia (via $St$) or gravity (via $Sv$). In the next section, we will attempt to clarify those effects. 

\subsection{Fixed $Sv$, Varying $St$}

\begin{figure}[h]
    \centering
    \includegraphics[width=0.95\textwidth]{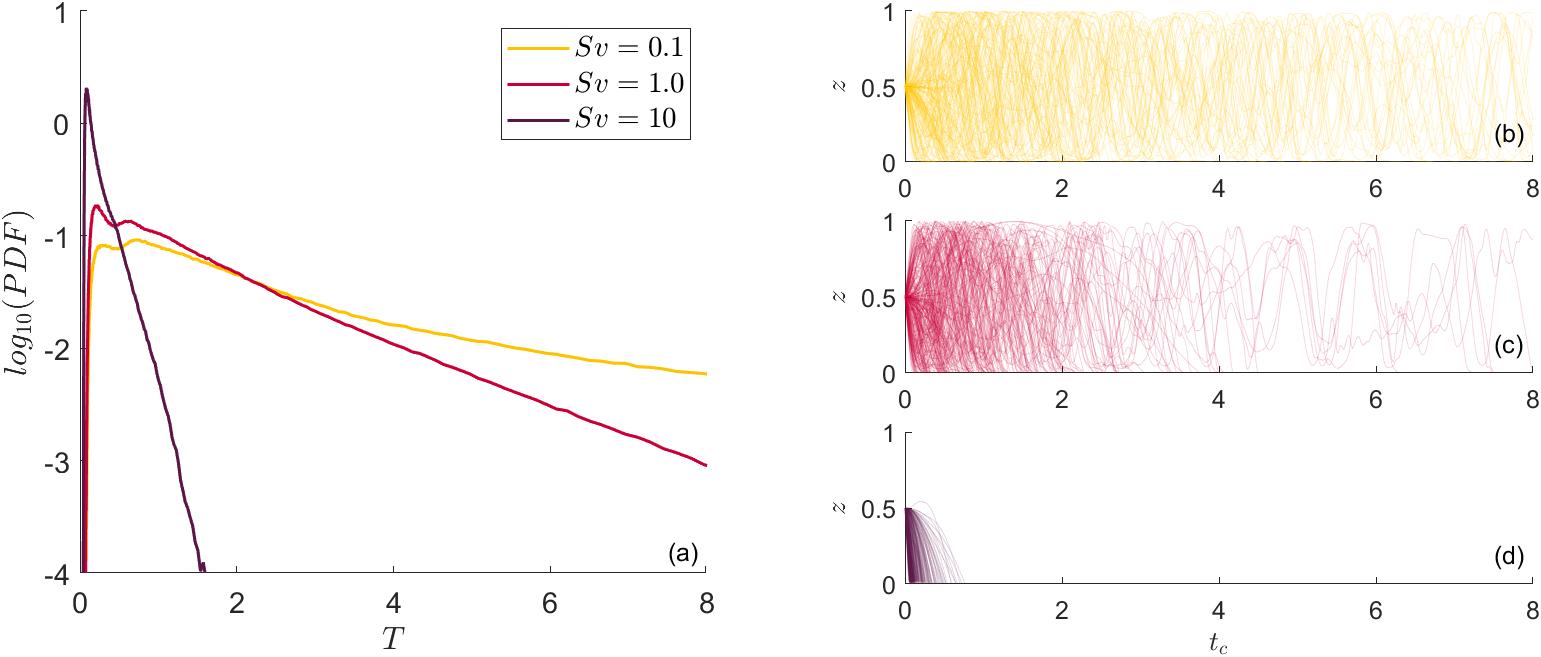}
    \caption{DNS results from the constant $Sv$ cases, for all particles over the range $St = [10^{-3},10^{2}]$: residence PDFs (a) and representative trajectories (b-d)}
    \label{fig:uncoupled_dns1}
\end{figure}

For this portion of the analysis, we choose three constant values of $Sv = [0.1, 1.0, 10]$. In the previous case, we were limited in how high of an $St$ value we could consider because a corresponding $Sv$ greater than $10^2$ results in the particle falling out almost immediately. Here, however, we shift the range of values to $St = [10^{-3},10^{2}]$ to encompass the transition between low to substantial inertial effects. Figure \ref{fig:uncoupled_dns1} shows the resulting residence PDFs along with some representative trajectories measured in the DNS. 

\begin{figure}[t]
    \centering
    \includegraphics[width=0.72\textwidth]{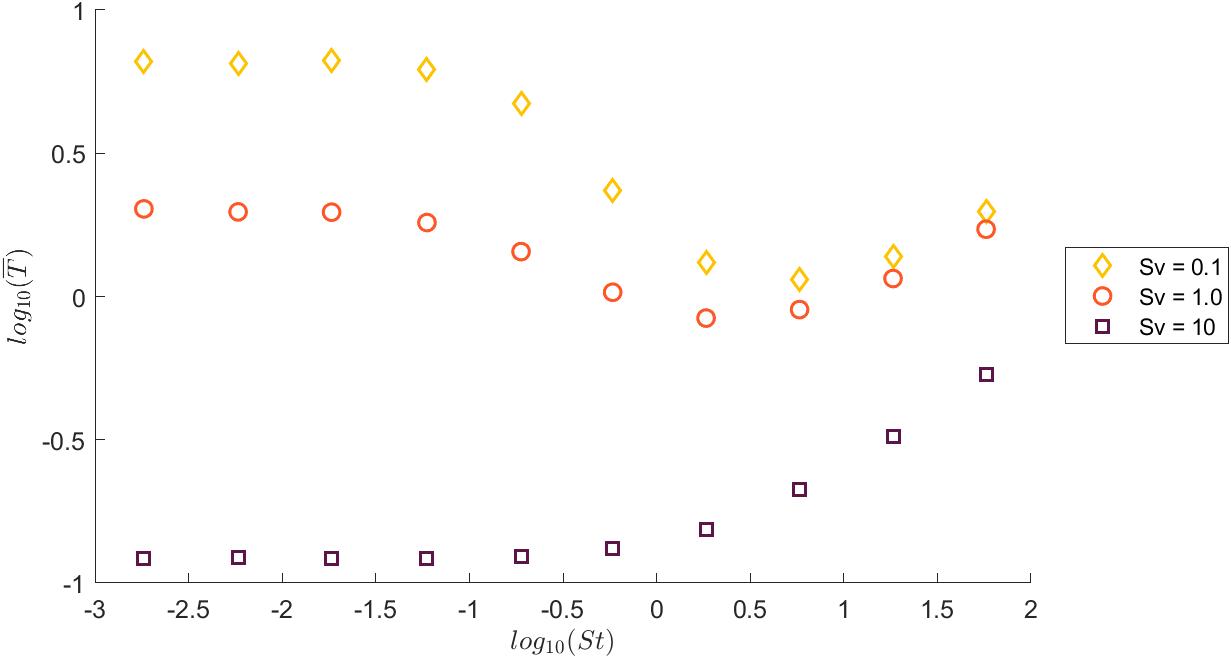}
    \caption{Summary of the mean non-dimensional residence times ($T$) over the range $St = [10^{-3},10^2]$ for each constant $Sv$}
    \label{fig:uncoupled_dns2}
\end{figure}

As expected, it can be seen that for the same range of Stokes numbers, increasing the settling velocity decreases the residence time on average. Also of note is that by $Sv = 10$, none of the particles are able to complete an elevator trip as they fall immediately out of the domain owing to their large settling velocity. Of more interest, however, is how the residence times vary as a function of $St$ given a constant $Sv$. These results are presented in Figure \ref{fig:uncoupled_dns2}, which shows that for low Stokes numbers ($<10^{-1}$), the residence time remains solely a function of $Sv$. As $St$ increases, particle residence times initially decrease, but past $St \approx 10$ they begin to increase again.

\begin{figure}[t]
    \centering
    \includegraphics[width=0.98\textwidth]{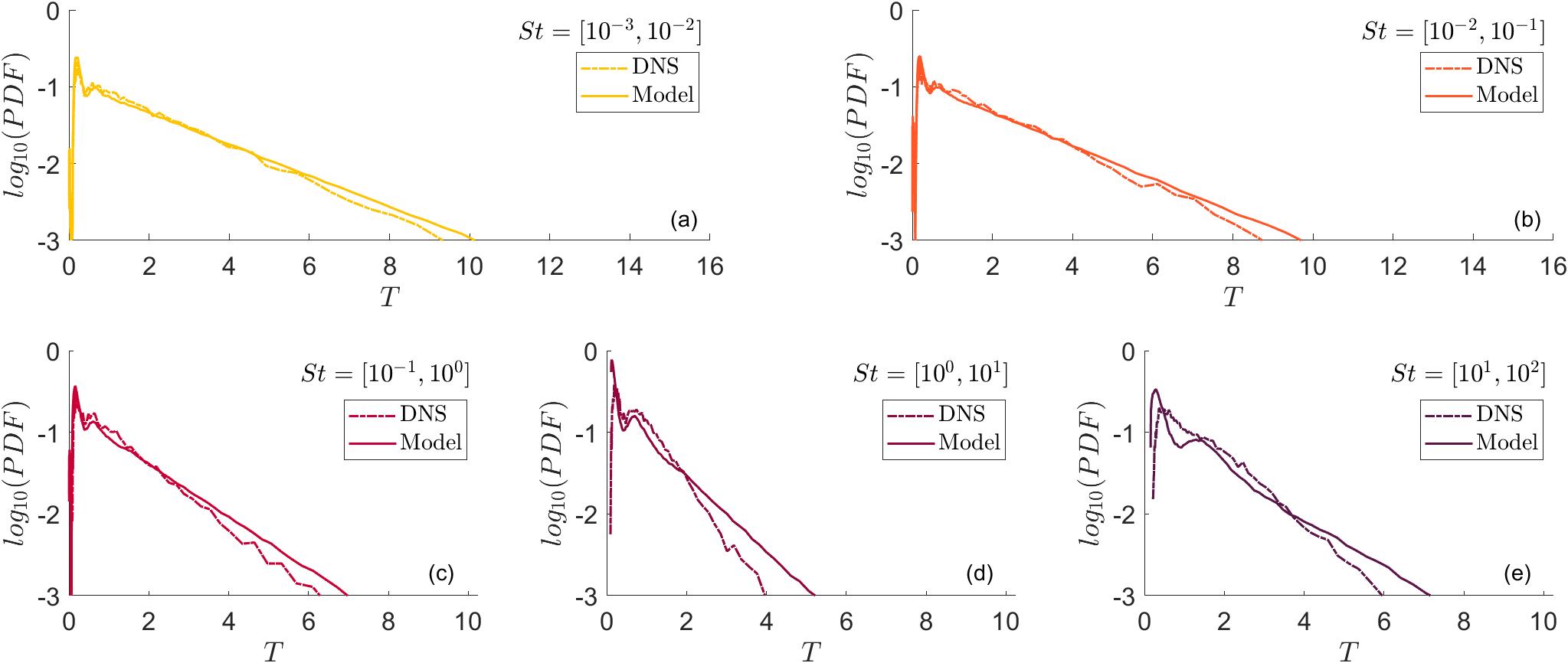}
    \vspace{-8pt}
    \caption{The results of the fixed $Sv$, varying $St$ cases as shown by the comparison of residence time distributions for the DNS and the model. For clarity, each order of magnitude of St is compared individually (a-e)}
    \label{fig:uncoupled_figure1}
\end{figure}

\begin{figure}[t]
    \centering
    \includegraphics[width=0.75\textwidth]{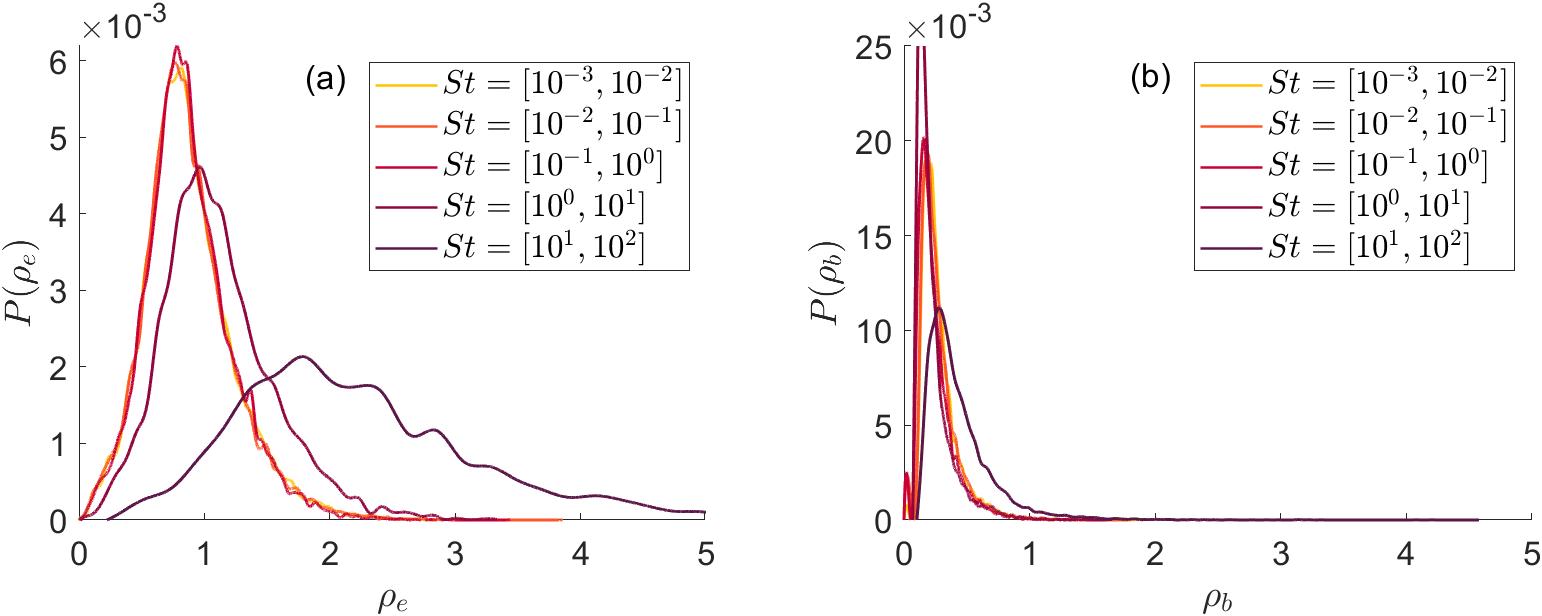}
    \vspace{-8pt}
    \caption{For the sample case of $Sv = 1.0$, distributions for the time to complete and elevator trip, $\rho_e$ (a) and the time to travel from the midplane to the bottom boundary, $\rho_b$ (b)}
    \label{fig:uncoupled_figure2}
\end{figure}

Figure \ref{fig:uncoupled_figure1} compares the PDFs for both the DNS and the model given the sample case of $Sv = 1.0$. We see that the model is again able to quantitatively replicate the PDF for  the entire range of $St$. In contrast to the case where $St$ and $Sv$ are linked, however, Figure \ref{fig:uncoupled_figure2} shows that at a fixed settling velocity, $St$ can change the distributions of $\rho_e$ and $\rho_b$. For the largest particles ($St > 10$), we see an increase in the mean residence time in Figure \ref{fig:uncoupled_dns2}. The two main causes for this can be found in Figure \ref{fig:uncoupled_figure2}. First, we see that the time to complete an elevator trip ($\rho_{e}$) increases substantially, due to their delayed response to turbulence combined with the tendency to filter out small-scale motions. Secondly, the higher inertia also implies that they take a longer to approach their terminal velocity, as evidenced by the increased time to travel from the midplane to the bottom boundary ($\rho_{b}$). 

\begin{figure}[t]
    \centering
    \includegraphics[width=0.95\textwidth]{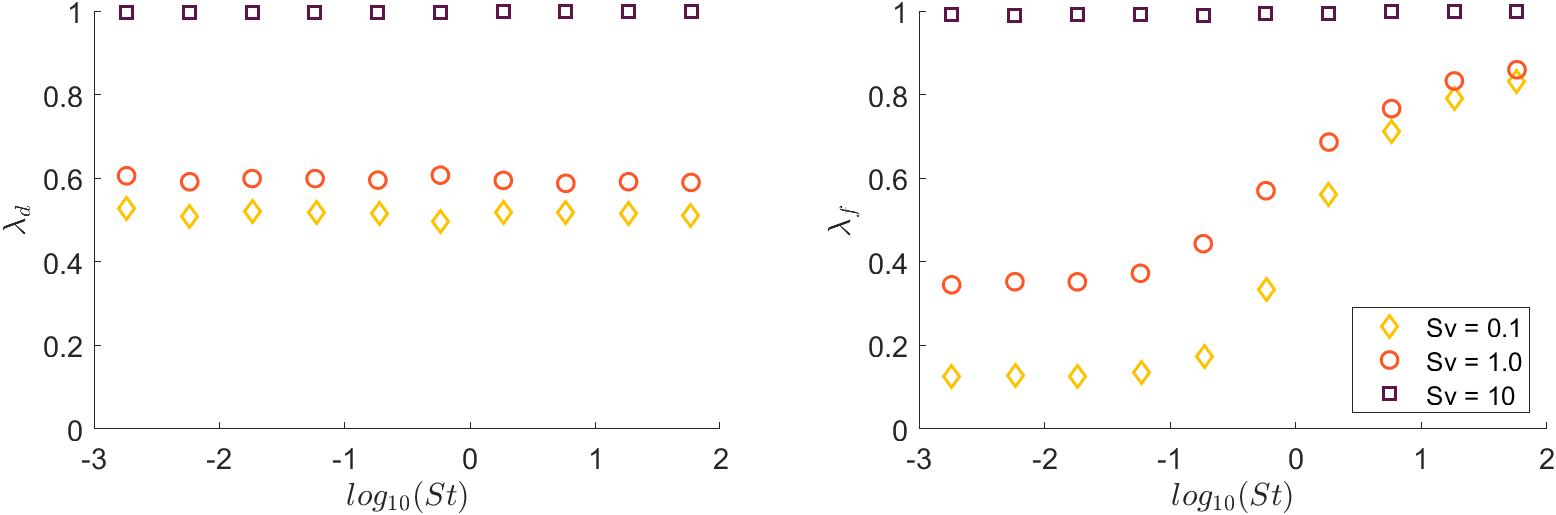}
    \vspace{-10pt}
    \caption{For all three constant $Sv$ cases, the percent chance of the particle initially having a downward velocity, $\lambda_d$ (a), and the percent chance of falling out after each elevator trip, $\lambda_f$ (b)}
    \label{fig:uncoupled_figure3}
\end{figure}

The other two model input parameters can be found in Figure \ref{fig:uncoupled_figure3} for all three constant $Sv$ values over the entire considered range of $St$. We can clearly see that $\lambda_d$ is solely a function of $Sv$. We also notice that $\lambda_f$ begins to increase once $St$ is greater than $10^{-1}$. This happens when particles are beginning to depart from streamlines, and are therefore flung towards the bottom boundary where they fall out. This would account for the initial dip in residence times that is seen in Figure \ref{fig:uncoupled_dns2}. At even higher $St$, the chance of falling out continues to approach 100\% as $St$ increases. This is because, as $St$ continues to increase, the drag force on the particle becomes negligible compared with that produced by gravity, which pulls it towards the bottom boundary.

\begin{figure}[t]
    \centering
    \includegraphics[width=0.75\textwidth]{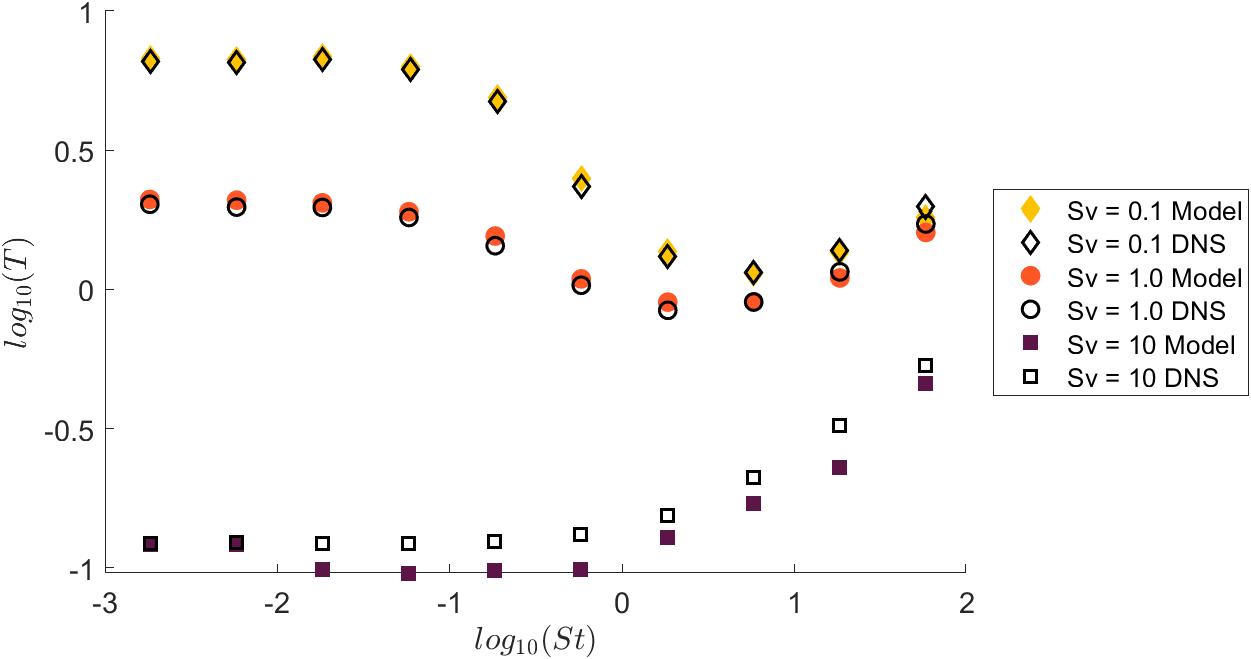}
    \vspace{-12pt}
    \caption{Validation of the model results, shown by the mean non-dimensional residence times ($T$) across the entire range of $St$ for each constant $Sv$ value}
    \label{fig:uncoupled_figure4}
\end{figure}

Figures \ref{fig:uncoupled_figure2} and \ref{fig:uncoupled_figure3} can be summarized in the following way. For small $St$ ($<10^{-1}$), all particles are subject to the same, turbulence-based convective time scale during their elevator trips, and the percent chance of falling out remains constant, leading to little change in their overall residence times. As $St$ begins to increase ($10^{-1}<St<10^1$), elevator trips are still governed by the flow convective time scale, but the percent chance of falling out begins to increase due to the particles departing from streamlines, leading to a decrease in residence times. Once $St$ becomes very large ($>10^1$), the particles have enough inertia to strongly resist the effects of the flow, and the elevator trips themselves become longer since they are experiencing a low-pass-filtered version of the surrounding turbulence. Even in the $Sv = 10$ case where particles rarely complete elevator trips, their high inertia prevents them from reaching their Stokes terminal velocity. This results in an increase in residence times, and we would expect the residence times to continue to increase along with $St$. Figure \ref{fig:uncoupled_figure4} shows how the model is able to match this behavior that was already demonstrated in the DNS results.

\clearpage

\section{Discussion and Conclusion}
In this work, we have proposed a stochastic model that reduces the complexities of particle-laden turbulent Rayleigh-B\'{e}nard flow to a simple conceptual picture. Motivated by the Pi Chamber experimental facility \citep{ChangPi}, we used one-way coupled DNS with Lagrangian particles to model their behavior and record statistics associated with their residence times. We focused in particular on the independent roles of $St$ and $Sv$ in dictating particle residence times in the flow.

In order to simplify the complex motion of the particles in the flow, we introduced the idea of an `elevator trip' which is the approximately sinusoidal motion generated by the convective B\'{e}nard cells. The four important statistics to describe this motion are: the chance of a particle having an initial downward velocity, the time it takes to complete an elevator trip, the chance of falling out after an elevator trip, and the time it took to fall from the midplane to the bottom boundary. We have demonstrated that when these input statistics for the model are prescribed using DNS data, then the model predictions for the residence times accurately replicate the DNS results. That it should do so is not at all trivial given the complexity of particle motion in turbulent Rayleigh-B\'{e}nard flow, and the simplicity of the approximations underlying the model. This test accomplishes two things. First, it demonstrates that with perfect knowledge of the inputs, the stochastic model provides a very good approximation of both the mean and full distribution of residence times. Secondly, it shows that this simple conceptual framework provides insight into the physical phenomena governing the particle residence times in this system.

When $St$ and $Sv$ are coupled, as they would in experimental conditions, we saw that the amount of time to complete an elevator trip remained constant, but as $St$ increased more particles had an initial downward velocity, and they were more likely to fall out after a given elevator trip. This unsurprisingly results in larger particles having shorter residence times. To clarify the independent roles of $St$ and $Sv$, we chose three constant, representative values of $Sv$ and varied $St$ for each to isolate the effects of inertia. In these runs, we saw that for small $St$, residence times are solely a function of $Sv$ since the lifetime is ultimately dictated by the particles settling through the boundary layer, which is not aided by particle inertia when $St$ is small. However as $St$ begins to increase, so does the chance of falling out after each elevator trip, leading to an initial decrease in residence times. This corresponds to particles departing from streamlines and being flung out of the turbulent core of the domain. For the largest $St$, the increased inertia leads to longer elevator trips and slow relaxation to their Stokes terminal velocity, resulting in a reversion towards longer residence times. Using this knowledge, we can look at our coupled $St$ and $Sv$ results in a new light. For particles with $St < 10^{-4}$, motion is dominated by the number of elevator trips, differing only slightly due to the amount of time required to settle through the boundary layer. 

In the end, we demonstrated that the simple conceptual framework underlying the stochastic model provides a helpful way to understand the behavior of the particles in the flow, and if the input statistics are perfectly described, then it also provides accurate approximations for both the mean residence times and their complete probability distributions. At the moment, the model relies on DNS data to prescribe the input statistics. In future work, a key point will be to develop models for the four statistical inputs themselves, so that the stochastic model for the particle residence times is fully closed. From the results we can already see that the convective time scale $(t_c)$ is helpful in predicting the average time it takes to complete an elevator trip. While it is beyond the scope of this paper, there is promise in finding similar relationships for the other inputs. Figure \ref{fig:uncoupled_figure3} shows that the chance of an initial downward velocity is a function of $Sv$. Since we have chosen to initialize the particles with zero velocity at the midplane where the fluid has no mean vertical velocity, $\lambda_d$ approaches 50\% as $Sv \rightarrow 0$. Additionally, since the length of each elevator trip is decided independently of the others, we are free to vary particle properties after each trip. If the particle size were to be updated based on the relative humidity encountered, this model could potentially be extended to evaporating particles, with even more direct applications to the Pi Chamber.

\begin{acknowledgements}
This research was supported by NSF grant AGS-2227012. The authors gratefully acknowledge computing support from the Notre Dame Center for Research Computing.

\end{acknowledgements}

\clearpage



\bibliography{main}

\end{document}